\documentclass[pdflatex,sn-mathphys-num]{sn-jnl}% Math and Physical Sciences Numbered Reference Style
\usepackage{graphicx}%
\usepackage{multirow}%
\usepackage{amsmath,amssymb,amsfonts}%
\usepackage{amsthm}%
\usepackage{mathrsfs}%
\usepackage[title]{appendix}%
\usepackage{xcolor}%
\usepackage{textcomp}%
\usepackage{manyfoot}%
\usepackage{booktabs}%
\usepackage{algorithm}%
\usepackage{algorithmicx}%
\usepackage{algpseudocode}%
\usepackage{listings}%
\usepackage[table]{xcolor}
\newcommand{\re}[1]{\textcolor{black}{#1}} % revise

\newcommand{\eg}[1]{\textit{e.g.}, #1}
\usepackage{array}
\theoremstyle{thmstyleone}%
\theoremstyle{thmstyletwo}%

\theoremstyle{thmstylethree}%

\begin{document}

\title[Artificial intelligence as a surrogate brain: Bridging neural dynamical models and data]{Artificial intelligence as a surrogate brain: Bridging neural dynamical models and data}


\author{Yinuo Zhang$^{1,\dagger}$}
\author{Demao Liu$^{2,3,4,\dagger}$}
\author{Zhichao Liang$^{1,\dagger}$}
\author{Jiani Cheng$^{2,3,4}$}
\author{Kexin Lou$^{1}$}
\author{Jinqiao Duan$^{5,4,6}$}
\author{Ting Gao$^{2,3,4,6,*}$}
\author{Bin Hu$^{7,*}$}
\author{Quanying Liu$^{1,*}$}

\affil{$^1$Department of Biomedical Engineering, Southern University of Science and Technology, Shenzhen, Guangdong, 518055, China}
\affil{$^2$School of Mathematics and Statistics, Huazhong University of Science and Technology, China}
\affil{$^3$Center for Mathematical Science, Huazhong University of Science and Technology, China}
\affil{$^4$Steklov-Wuhan Institute for Mathematical Exploration, Huazhong University of Science and Technology, China}
\affil{$^5$Department of Mathematics and Department of Physics, Great Bay University, Dongguan, China}
\affil{$^6$Guangdong Provincial Key Laboratory of Mathematical and Neural Dynamical Systems, Great Bay University, China}
\affil{$^7$School of Medical Technology, Beijing Institute of Technology, Beijing, 100081, China}

\affil{\textbf{Corresponding authors.} Email: liuqy@sustech.edu.cn; bh@bit.edu.cn; tgao0716@hust.edu.cn}

\affil{Equally contributed to this work}


\abstract{Recent breakthroughs in artificial intelligence (AI) are reshaping the way we construct computational counterparts of the brain, giving rise to a new class of ``surrogate brains''. In contrast to conventional hypothesis-driven biophysical models, the AI-based surrogate brain encompasses a broad spectrum of data-driven approaches to solve the inverse problem, with the primary objective of accurately predicting future whole-brain dynamics with historical data. Here, we introduce a unified framework of constructing an AI-based surrogate brain that integrates forward modeling, inverse problem solving, and model evaluation. Leveraging the expressive power of AI models and large-scale brain data, surrogate brains open a new window for decoding neural systems and forecasting complex dynamics with high dimensionality, nonlinearity, and adaptability. We highlight that the learned surrogate brain serves as a simulation platform for dynamical systems analysis, virtual perturbation, and model-guided neurostimulation. We envision that the AI-based surrogate brain will provide a functional bridge between theoretical neuroscience and translational neuroengineering.}




\maketitle
\section{INTRODUCTION}\label{sec1}

The quest to understand and predict brain activity has long relied on mathematical and computational models of neural dynamics,
which \re{provide} a rigorous mathematical foundation to describe how neural states evolve in time and space~\cite{siettos2016multiscale}. However, despite decades of progress with mechanistic models, ranging from single neuron Hodgkin–Huxley equations~\cite{hodgkin1952currents} to large-scale neural mass models~\cite{breakspear2017dynamic}, traditional forward modeling remains fundamentally constrained by fixed-form equations and population-averaged parameters. These constraints limit its ability to capture the rich, nonlinear, and context-dependent dynamics of individual brains, particularly in high-dimensional, noisy, and behaviorally relevant settings.

Recent advances in machine learning, coupled with the explosive growth of large-scale neural datasets~\cite{elam2021human, miller2016multimodal, kiessner2023extended}, \re{hold} promising potential to model the spatio-temporal dynamics of the brain in a data-driven fashion~\cite{di2023unraveling}. AI models such as recurrent neural network (RNNs)~\cite{durstewitz2023reconstructing} and transformer-based architectures~\cite{achiam2023gpt} have achieved state-of-the-art performance in forward modeling across domains. This convergence has catalyzed the emergence of surrogate brains: a broad family of computational models—spanning white-box formulations with explicit mechanistic equations~\cite{wang2023delineating, jirsa2023personalised}, black-box deep networks trained directly from data~\cite{luo2025mapping, pandarinath2018inferring}, and gray-box hybrids that embed neurobiological priors into adaptive architectures~\cite{brenner2022tractable, zeng2023braincog}. 
Unlike traditional modeling approaches that primarily emphasize anatomical correspondence, surrogate brains place priority on dynamical fidelity and predictive utility, 
offering a flexible computational counterpart to living brains that integrates prior knowledge with scalable, data-driven learning.

A surrogate brain serves not only as a forward model, predicting future neural states and transitions between brain states, but also a solver of the inverse problem: inferring latent states and dynamical rules that best explain observed neural activity. This dual role is essential for bridging theory and experiment, enabling mechanistic interpretation of complex signals, and supporting applications such as hypothesis testing, virtual perturbation, and model-guided neurostimulation.
Traditionally, inverse problems in neuroscience have been approached by fitting predefined dynamical models to data through optimization~\cite{Brambilla2017ModelDrivenSE} or Bayesian inference~\cite{tolley2024methods}. While these methods have yielded valuable insights, their reliance on certain equation structures limits their expressivity and adaptability. In contrast, surrogate brains can directly learn \re{dynamical} rules from data, supporting white-box, black-box, and gray-box formulations that adaptively balance mechanistic interpretability with empirical fidelity.

\begin{figure}[htbp]
    \centering
    \includegraphics[width=\textwidth]{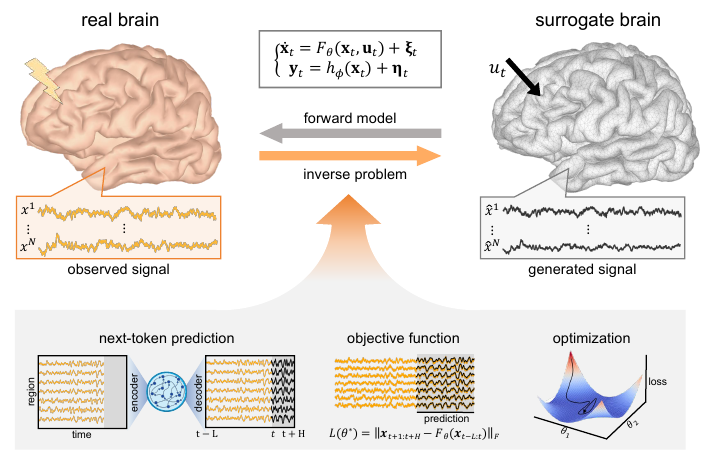}
    \caption{Conceptual framework for AI-based surrogate brains in neural dynamical systems. 
    A surrogate brain is constructed through two interconnected processes: forward modeling and inverse problem solving. 
    \textit{Forward modeling} describes how latent brain states $\mathbf{x}_t$ evolve according to a dynamical operator $F_\theta(\cdot)$ and produce observable signals via an observation mapping $h_\phi(\cdot)$. Here, $\mathbf{u}_t$ represents external inputs, $\boldsymbol{\xi}_t$ captures intrinsic dynamical noise, and $\boldsymbol{\eta}_t$ models measurement uncertainty. 
    \textit{Inverse problem solving} entails learning $F_\theta(\cdot)$ and $h_\phi(\cdot)$ from data, which involves selecting training strategies (\eg{next-token prediction}), defining objectives, and optimizing parameters. The framework accommodates white-box, gray-box, and black-box formulations of $F_\theta(\cdot)$ and $h_\phi(\cdot)$, enabling the surrogate brain to flexibly integrate mechanistic priors with data-driven adaptation. Together, these steps yield a personalized, predictive model of brain dynamics that supports mechanistic insight, virtual experimentation, and model-guided neurostimulation.
    }
    \label{fig:framework}
\end{figure}

% \begin{figure*}[htbp]
%   \noindent
%   \hspace*{-0.13\paperwidth}% Shift the entire minipage left
%   \begin{minipage}{0.78\paperwidth}% Minipage spans the full page width
%     \includegraphics[width=0.78\paperwidth]{Figures/Figure_R1/R1_figure1.pdf}%
%     \caption{Conceptual framework for AI-based surrogate brains in neural dynamical systems. 
%     A surrogate brain is constructed through two interconnected processes: forward modeling and inverse problem solving. 
%     \textit{Forward modeling} describes how latent brain states $\mathbf{x}_t$ evolve according to a dynamical operator $F_\theta(\cdot)$ and produce observable signals via an observation mapping $h_\phi(\cdot)$. Here, $\mathbf{u}_t$ represents external inputs, $\boldsymbol{\xi}_t$ captures intrinsic dynamical noise, and $\boldsymbol{\eta}_t$ models measurement uncertainty. 
%     \textit{Inverse problem solving} entails learning $F_\theta(\cdot)$ and $h_\phi(\cdot)$ from data, which involves selecting training strategies (\eg{next-token prediction}), defining objectives, and optimizing parameters. The framework accommodates white-box, gray-box, and black-box formulations of $F_\theta(\cdot)$ and $h_\phi(\cdot)$, enabling the surrogate brain to flexibly integrate mechanistic priors with data-driven adaptation. Together, these steps yield a personalized, predictive model of brain dynamics that supports mechanistic insight, virtual experimentation, and model-guided neurostimulation.}
%     \label{fig:framework}
%   \end{minipage}
% \end{figure*}

In this review, we focus on data-driven approaches for solving inverse problems in neural dynamical systems. We propose a surrogate brain framework that \re{bridges} the neural dynamical model with observed neural data (Fig.~\ref{fig:framework}). The framework covers: (i) modeling neural dynamics using varying degrees of prior knowledge, (ii) solving the inverse problem and learning model parameters via fitting data, and (iii) evaluating the learned surrogate brain on signal fidelity, functional consistency, and task-level performance. 
Building on this foundation, we highlight their applications in system analysis, virtual experimentation, and neural stimulation.
Finally, we discuss the remaining challenges and future directions in realizing fully personalized, real-time surrogate brains.

\section{MODELING OF NEURAL DYNAMICAL SYSTEMS}\label{sec2}
To address the inverse problem, the first step is to construct an appropriate forward model, which simulates the brain's neural activity over time. The forward model is defined by the following equations:
\begin{align}
    \dot{\mathbf{x}}_t &= F_{\theta}(\mathbf{x}_t, \mathbf{u}_t) + \boldsymbol{\xi}_t, \label{eq:general}\\
    \mathbf{y}_t &= h_{\phi}(\mathbf{x}_t) + \boldsymbol{\eta}_t. \notag
\end{align}
Here, $\mathbf{x_t}$ represents the latent brain state at time $t$, which encapsulates the internal neural dynamics, while $\mathbf{u_t}$ denotes the external inputs to the system (\eg{sensory stimuli or experimental interventions}). The term $F_\theta(\cdot)$ describes the system dynamics, parameterized by $\theta$, which governs how the brain's state evolves over time. Additionally, $\boldsymbol{\xi}_t$ represents the process noise, which captures any unmodeled dynamics or inherent randomness in the system's evolution. The second equation defines the observation model, where $\mathbf{y}_t$ is the observable signal at time $t$ (\eg{EEG or fMRI}). This observable signal is related to the latent brain state $\mathbf{x}_t$ through the function $h_\phi(\cdot)$, which maps the brain's internal state to the observed data, with $\boldsymbol{\eta}_t$ representing measurement noise or uncertainty in the observations. 
This formulation, as shown in Eq~\ref{eq:general}, serves as the foundation for the surrogate brain framework in Fig.~\ref{fig:framework}. The forward model generates neural dynamics over time, based on parameters $\theta$, initial conditions $\mathbf{x_0}$, and external inputs ${\mathbf{u}(0), \mathbf{u}(1), \dots, \mathbf{u}(T)}$. The framework uses data-driven learning to estimate $F_\theta(\cdot)$ and $h_\phi(\cdot)$ from real neural data, enabling personalized, predictive models of brain activity.
In this review, we do not consider the physical mapping from the observation space to the state space. Instead, we simplify the objective function Eq~\ref{eq:general} to the identity mapping, $\mathbf{y}_t = \mathbf{x}_t$. As a result, the system can be described as ${\dot{\mathbf{x}}_t} = F_\theta(\mathbf{x}_t, \mathbf{u}_t) + \boldsymbol{\xi}_t$.

\begin{figure}[htbp]
    \centering
    \includegraphics[width=\textwidth]{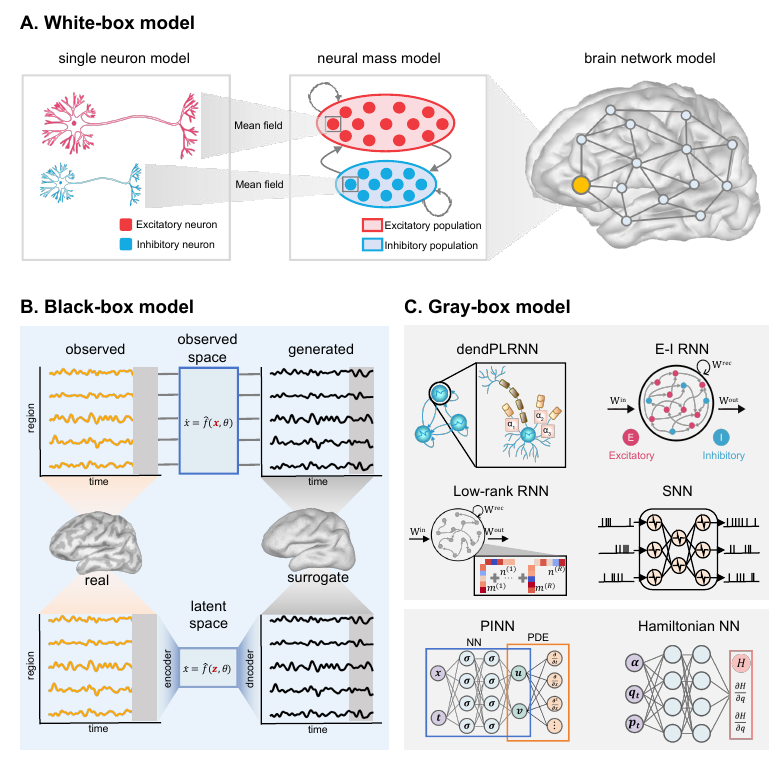}
    \caption{Three types of neural dynamical models:  white-box, black-box, and gray-box models. 
    A. White-box models at multiple scales.   
    Left: biophysical single-neuron models; center: a neural-mass model obtained by mean-field reduction of many single neurons; right: a whole-brain network in which each node is a neural-mass model.
    \re{\emph{Neural mass model in panel A is an original schematic adapted from the conceptual framework in Ref.~\cite{breakspear2017dynamic}.}}
    B. Black-box models.
    Top: modelling directly in the observation space, where a neural network predict the future neural dynamics, $\mathbf{\dot{x}}=\hat{f}(\mathbf{x},\theta)$; bottom: modelling in a latent space, where the data are first encoded/decoded and dynamics are learned as $\dot{\mathbf{z}}=\hat{f}(\mathbf{z},\theta)$.  
    \textbf{C. Gray-box models incorporating priors.} 
    Top: neural dynamics models constrained by neuroscientific priors (including dend-PLRNN,~\re{\emph{reproduced from} Ref.~\cite{brenner2022tractable} under a CC BY 4.0 license)}, E-I RNN~\cite{song2016training}, Low-rank RNN~\cite{pellegrino2023low} and Spiking Neural Network (SNN)); bottom: dynamical models constrained by physical-law priors(including Physics-Informed Neural Network (PINN) and Hamiltonian Neural Network (HNN)). 
    }
    \label{fig:modeling}
\end{figure}

From a machine learning perspective, we categorized neural dynamical models into three classes (Fig.~\ref{fig:modeling}), namely \textbf{white-box models} (model-driven), \textbf{black-box models} (data-driven), and \textbf{gray-box models} (hybrid approach), depending on the incorporation of prior knowledge into the model structure. Specifically, white-box models, such as the Hodgkin-Huxley model, rely heavily on detailed neurobiological principles and physical laws to simulate specific neural processes. In contrast, black-box models, such as artificial neural networks, are primarily data-driven, making minimal use of prior knowledge but excelling in their ability to fit complex \re{real-world} data. Gray-box models aim to bridge the gap between the two, integrating prior knowledge with data to enhance interpretability while maintaining strong predictive performance.

\subsection{White-box models}
White-box models are distinguished by their capacity to incorporate mechanistic principles derived from neurobiology and physics, providing a structured and interpretable approach for simulating neural dynamics across hierarchical levels of organization, from single neurons to large-scale brain networks.

A unified dynamical systems framework \re{for} a single neuron can be \re{expressed} as:
\begin{equation}
    C\frac{dv}{dt}=-\sum_{k}I_{k}(t)+I_{ext}(t).
\end{equation}
where $C$ represents the membrane capacitance, $\frac{dv}{dt}$ is the rate of change of the membrane potential, $\sum_{k}I_{k}(t)$ represents the sum of ionic and synaptic currents, and $I_{\text{ext}}(t)$ is the externally applied current.

Classical models like the Hodgkin-Huxley (HH) equations \re{are} viewed as biophysically enriched specializations of this framework~\cite{hodgkin1952currents}.
The HH model uses a set of nonlinear differential equations to characterize ion fluxes across the neuronal membrane, thereby generating action potentials and enabling signal propagation~\cite{goldman2008new}. 
While providing detailed and mechanistically interpretable insights into neuronal firing dynamics~\cite{he2024network}, its computational complexity poses challenges for large-scale simulations.
To address this, simplified models such as the Integrate-and-Fire (IF)~\cite{lapicque1907recherches} and Leaky Integrate-and-Fire (LIF) models~\cite{liu2001spike} are commonly employed, \re{especially} for large-scale simulations of spiking neural networks. 
While these models are less detailed than the HH model, they provide a balance between simplicity and computational efficiency, making them well-suited for large-scale simulations of neural networks~\cite{yao2022glif}.

At the neural population level, white-box models extend the principles of single-neuron dynamics to capture the interactions within and between neural populations. 
This transition is typically achieved via the mean-field approximation, which replaces detailed, neuron-specific dynamics with macroscopic variables(\eg{mean membrane potential}), reflecting the collective activity of the population~\cite{deco2008dynamic}. 
Neural Mass Models (NMMs), a prominent class of such models, build on mean field approximation to describe population-level dynamics using low-dimensional, biophysically interpretable equations~\cite{cooray2023global}. By reducing the complexity of high dimensional neuronal systems, NMMs provide a computationally efficient framework for simulating large-scale neural activity. They are typically formulated as systems of ordinary differential equations (ODEs) that represent the temporal evolution of population states, such as the membrane potential of excitatory neurons.
Notable examples of NMMs include the Wilson-Cowan model~\cite{wilson1972excitatory} and the Jansen-Rit model~\cite{jansen1993neurophysiologically}, which represent the dynamics of excitatory and inhibitory populations using coupled ODEs. Other notable NMMs include the Epileptor~\cite{jirsa2014nature}, the Thalamus model~\cite{schellenberger2016thalamocortical}, and the Larter-Breakspear model~\cite{larter1999coupled}. 
A merit of NMMs \re{is incorporating} neurobiological factors, such as synaptic connectivity, neurotransmitter dynamics, and network topology, to simulate phenomena like neural oscillations (\eg{EEG~\cite{sotero2007realistically}} ), synchronized activity (\eg{visual attention~\cite{chen2022spatiotemporal}}), and transitions between different brain states (\eg{epileptic seizure onset~\cite{jirsa2014nature}}). Another merit of NMMs is their computational efficiency, \re{by simplifying and reducing} the complexity of large-scale network simulations while maintaining the essential dynamics of individual neural interactions~\cite{breakspear2017dynamic}.

While NMMs primarily focus on temporal dynamics, they can be generalized to incorporate spatial effects through neural field models (NFMs)~\cite{jirsa1997derivation}. These models extend NMMs by describing the spatial propagation of neural activity alongside its temporal evolution. NFMs enable the study of phenomena like traveling waves~\cite{inglis2016general}, synchronization patterns~\cite{lv2019synchronization}, and multi-scale spatial interactions in the brain~\cite{sanz2018nftsim}. NFMs offer a finer grained depiction of neural activity across space, for instance, Jirsa et al. \re{demonstrated} that NFMs can achieve up to 1000-fold improvements in spatial resolution over NMMs when localizing epileptogenic networks~\cite{wang2023delineating}. More recently, neural field theory has been used to investigate how brain geometry constrains activity patterns~\cite{pang2023geometric}, with successful applications to source localization tasks~\cite{wang2024advancing}.
By bridging detailed local interactions and global brain activity, NMMs and NFMs provide a critical framework for understanding population level neural dynamics within a biologically grounded, mathematically tractable paradigm.

At the large-scale brain network level, white-box models aim to capture the macroscopic organization of brain activity by modeling interactions between distinct brain regions and functional networks. These models often employ coupled differential equations or neural field theories to simulate dynamic interactions across brain regions. A prominent approach is the Brain Network Models (BNMs), which integrates NMM as \re{the} node dynamics to simulate the interactions and dynamics of entire brain networks~\cite{honey2007network, ghosh2008noise}. 

\begin{equation}
\begin{split}
  \frac{d}{d t} x_i(t) &= f\left(x_i(t)\right) + \sum_{j=1}^N g\left(G_{i j}, x_i(t), x_j\left(t-D_{i j}\right)\right) + \xi_i(t).
\label{Jirsa}
\end{split}
\end{equation}
where $x_i$ represents the state of the $i^{th}$ brain region, and the local node dynamics $f$ of the node $i$ are determined by NMM. The second term describes the global coupling between interconnected regions, determined by the adjacency matrix $G$, which typically encodes structural connectivity derived from empirical data (\eg{Diffusion Tensor Imaging (DTI)}). This term also incorporates the delayed influence of the $j$-th brain region on the $i$-th, \re{which is} defined by the delay matrix $D$. 
The third term, $\xi_i(t)$, accounts for noise input to each node.
This allows BNMs to simulate how both structural and functional interactions contribute to large-scale brain activity~\cite{deco2013resting}. 
By integrating whole brain structural and functional information, BNMs open a window to understanding how distributed neural circuits coordinate to support complex cognitive functions, such as cognition, perception, and emotion~\cite{wang2024virtual}.

In support of these white-box models, several public computational platforms have been developed to implement BNMs, including BrainPy~\cite{wang2023brainpy}, The Virtual Brain (TVB)~\cite{Hashemi2020TheBV}, and neurolib~\cite{cakan2023neurolib}. 
Notably, when coupling local dynamical models into a whole brain framework, the required number of parameters increases exponentially. Thus, it is necessary to develop more advanced inverse problem solving approaches for parameter inference~\cite{zhang2024framework,lu2024simulation}.

\subsection{Black-box models}

Black-box models provide a data-driven alternative to white-box approaches for representing complex brain dynamics, especially when the underlying biophysical mechanisms are poorly understood or cannot be expressed explicitly with differential equations. In contrast to mechanistic models, black-box approaches make no explicit assumptions and instead learn neural dynamics implicitly through ANNs. As universal approximators, ANNs can theoretically capture any data pattern~\cite{di2023unraveling}, making them particularly suited for modeling the nonlinear and high dimensional processes characteristic of brain activity. Over the past decade, AI-based black-box models, such as RNNs, Neural ODEs~\cite{chen2018neural}, MAMBA~\cite{gu2023mamba} and transformers~\cite{abibullaev2023deep}, have demonstrated remarkable success in modeling dynamic systems and predicting complex data across various domains, from weather forecasting~\cite{rasp2020weatherbench} to molecular dynamics~\cite{schutt2017schnet}, and from financial modeling~\cite{liu2023fingpt} to large-scale language modeling~\cite{achiam2023gpt, gu2023mamba}. These advances now enable their direct application to neural dynamical systems, where black-box models can capture the intricate features of brain activity (Fig.~\ref{fig:modeling}\textbf{B}).%Typically, black-box models are initialized with random parameters and optimized by fitting empirical neural data, such as EEG, fMRI, or calcium imaging. This process is referred to as data assimilation in mathematics, model training in machine learning, or inverse problem-solving in this context. 

The success of black-box models depends heavily on aligning the model architecture with the specific characteristics of the neural signals being analyzed. Different modalities, such as fMRI and EEG, often require distinct AI architectures due to their unique temporal and spatial properties. For fMRI, Luo et al. showed that a simple multilayer perceptron (MLP) can predict future dynamics from the preceding three time steps within a self-supervised framework. Despite its simplicity, the model accurately \re{forecasts} brain activity and revealed latent structural relationships across regions~\cite{luo2025mapping}. In contrast, EEG demands architectures that can handle rapid temporal fluctuations~\cite{yang2023perturbing}. Pankka et al. applied WaveNet, a probabilistic deep learning model, to predict signals across multiple frequency bands. Their model captured amplitude, phase, and oscillatory peaks in theta and alpha activity, \re{demonstrated} its ability to represent fine-grained temporal patterns~\cite{pankka2024forecasting}. These studies underscore the importance of tailoring AI model architectures to the properties of specific neural signals, ensuring both accuracy and interpretability in modeling neural dynamics.

Transitioning from direct modeling in the observation space to modeling in latent spaces provides a complementary perspective on understanding neural dynamics. The motivation for latent space modeling stems from the inherent complexity of neural data, which is often high-dimensional, while the cognitive states or functional dynamics of interest typically reside in lower-dimensional subspaces. Latent space modeling aims to uncover low-dimensional representations through neural embeddings, enabling the study of their temporal evolution.
A prominent method in this domain is the variational autoencoder (VAE), which encodes high-dimensional data into a latent space, learns dynamic patterns within this reduced representation, and subsequently decodes the information back into the original space~\cite{kolter2019learning}. Building \re{upon} this framework, LFADS encodes neural spike trains into latent trajectories via an RNN, revealing low-dimensional patterns underlying complex dynamics~\cite{pandarinath2018inferring}. Similarly, Chen et al. combined Neural ODEs with VAEs to capture continuous latent dynamics~\cite{chen2018neural}. VAEs have proven effective in reconstructing low-dimensional trajectories. Beyond VAEs, RNNs have demonstrated \re{their} broad applicability in capturing latent neural dynamics. Leveraging their ability to model sequential data, RNNs have been employed in tasks such as simulating cerebellar computations~\cite{boven2023cerebro}, reverse-engineering cognitive functions~\cite{sohn2019bayesian, mastrogiuseppe2018linking}, and uncovering hierarchical dependencies across temporal and spatial scales in neural data~\cite{chen2021time, hu2019learning}. Manifold learning methods further complement these models by reducing dimensionality and improving interpretability~\cite{islam2023revealing, mao2024training}. While most latent \re{focus on} emphasize dimensionality reduction, an alternative strategy projects neural activity into higher-dimensional space. This facilitates the linearization of nonlinear dynamics, making them more amenable to prediction and control~\cite{kutz2016multiresolution}. For example, the Deep Koopman model maps epileptic brain dynamics into a high-dimensional linear framework, enabling online updates and model-based predictive control~\cite{liang2022online}.

Large pre-trained models based on Transformer architectures, such as GPT, have transformed data-driven modeling by learning universal representations from massive datasets~\cite{ouyang2022training, radford2021learning}. With the advancement of neural data collection technologies, large-scale, multimodal neural datasets such as HCP~\cite{elam2021human}, UK Biobank~\cite{miller2016multimodal}, TUAB~\cite{kiessner2023extended}, THINGS-EEG~\cite{grootswagers2022human}, HUP~\cite{kini2019virtual} and SEED~\cite{duan2013differential} have emerged, providing the foundation for neuroscience to adopt similar strategies. These datasets support the development of large-scale models that capture universal features of neural information~\cite{abibullaev2023deep, yi2024learning, ye2024neural, duan2024dewave}.
Examples include Brant~\cite{yuan2024brant, zhang2024brant}, a model with over 500 million parameters trained on intracranial EEG, and LaBraM~\cite{jiang2024large}, pre-trained on more than 2,500 hours of EEG recordings. A growing family of brain foundation models, such as NEURO-GPT~\cite{cui2024neuro}, NeuroLM~\cite{jiangneurolm}, and BrainWave~\cite{yuan2024brainwave}, highlights this trend.
In fMRI, contrastive alignment models, such as BrainCLIP~\cite{liu2023brainclip} and CLIP-MUSED~\cite{zhou2024clip} map fMRI embeddings into CLIP space to enable visual decoding, while self-supervised methods including BrainLM (MAE-style pretraining)~\cite{carobrainlm} and BrainMass~\cite{yang2024brainmass} focus on spatiotemporal dynamics and functional connectivity, respectively.
Despite these advances, a fundamental distinction remains. Pre-trained foundation models capture population level, task agnostic features, whereas a surrogate brain is an individualized dynamical state space model calibrated to a single person. Turning the former into the latter requires subject-specific adaptation through fine-tuning and calibration on the individual’s recordings, ensuring that the model’s forward dynamics reproduce that brain’s trajectories.

\subsection{Gray-box models}

White-box models offer mechanistic interpretability but often struggle to accommodate the complexity and variability of real neural data. In contrast, black-box models excel at learning from large-scale datasets yet typically lack transparency and biological plausibility. Positioned between these two extremes, gray-box models represent a hybrid modeling paradigm that integrates prior theoretical knowledge—derived from neuroscience or physics—with the flexibility of data-driven learning algorithms~\cite{ramezanian2022generative}. By embedding biologically or physically meaningful constraints into modern neural networks, these models preserve interpretability while retaining expressive power, offering a promising route toward constructing surrogate brains. Neuroscience priors serve as inductive biases that guide learning (Fig.~\ref{fig:modeling}\textbf{C}). They can shape model architectures directly or act as regularization terms during optimization, \re{as discussed further} in Section~\ref{sec3}.

One prominent example comes from the study of dendritic processing in neurons, which plays a pivotal role in the nonlinear integration of synaptic inputs. Inspired by this, researchers have developed artificial neurons and network modules that emulate dendritic processing~\cite{li2019dendritic, chavlis2025dendrites}. Liu et al. proposed Dit-CNN architectures augmented with quadratic neurons reflecting dendritic integration, which substantially boosted performance with minimal added complexity~\cite{liu2024dendritic}. Similarly, Durstewitz et al. introduced dend-PLRNN, incorporating dendritic mechanisms into recurrent dynamics to enhance nonlinear reconstruction while reducing model dimensionality~\cite{brenner2022tractable}.

Beyond individual neurons, broader organizational principles of the brain have also guided model development. The Dale principle, which states that neurons are either purely excitatory or inhibitory, has been incorporated into E-I RNN. As shown by Song et al., enforcing this biological fidelity stabilizes dynamics and structures temporal representations of cortical circuits~\cite{song2016training}. Some models, instead, emphasize the representation of higher-order assumptions about brain-wide activity. For example, the low-dimensional manifold hypothesis, suggesting that cortical activity resides on a constrained subspace has inspired the design of low-rank RNN. These models impose low-rank constraints on \re{the} recurrent weight matrices, yielding not only compact parameterizations but also interpretable latent dynamics~\cite{valente2022extracting, pellegrino2023low, costacurta2024structured}. 
To demonstrate the effectiveness of these prior constraints, we constructed an AI-based surrogate brain benchmark using RNNs, highlighting how dend-PLRNN~\cite{brenner2022tractable}, E-I RNN~\cite{song2016training}, and low-rank RNN~\cite{costacurta2024structured} exhibit distinct performance characteristics. \re{The related code is available on GitHub: \url{https://github.com/ncclab-sustech/Review_Surrogate_Brain.git}}, and further details can be found in the supplementary material.

Incorporating neuroanatomical features is another powerful approach in gray-box model development. For example, Dapello et al. introduced a specific layer into a hierarchical CNN to simulate the primate V1 area, which in turn allowed the model to better handle adversarial perturbations in image classification tasks~\cite{dapello2020simulating}. Jirsa et al. utilized DTI to create a data-driven whole-brain model, enabling the learning of the dynamics of the local nodes from both structural and functional data~\cite{sip2023characterization}. \re{The} integration of anatomical constraints allows these models to simulate more accurate whole-brain dynamics, offering region- and individual-specific parameterization for enhanced flexibility and precision.

Spiking Neural Networks (SNNs) provide another example of how neuroscientific priors can be embedded into model design. These networks exploit the pulse-based communication mechanism of biological neurons to process time-dependent information effectively~\cite{kasabov2014neucube}. 
Their architecture incorporates key neuronal mechanisms, including membrane potential accumulation, spike generation, and refractory periods, thereby closely mirroring biological dynamics. Learning in SNNs often relies on spike-timing-dependent plasticity (STDP), a rule inspired by Hebbian plasticity that adjusts synaptic strengths according to the relative timing of pre- and postsynaptic spikes~\cite{rachmuth2011biophysically}.
This intrinsic temporal dependence enables SNNs to encode and refine neural connections during learning, positioning them as a promising framework for neuromorphic computing. Several open-source platforms, such as BrainCog~\cite{zeng2023braincog}, SpikingJelly~\cite{fang2023spikingjelly}, and Nengo~\cite{bekolay2014nengo}, now support SNN simulation and application development, making these tools widely accessible to the research community.

While neuroscientific priors have been widely applied, physical principles offer a promising approach for augmenting gray-box models. Despite being less frequently employed in biological modeling, these physical laws present great potential to enhance both the performance and interpretability of gray-box methodologies (see Fig.\ref{fig:modeling} \textbf{C})). One example is PEMT-Net, which simulates the physical diffusion of neural signals using random walk processes, thereby capturing both local and long-range dependencies in fMRI signal decoding tasks. The results indicate that the physically enhanced model achieves a performance improvement exceeding 10 percentage points~\cite{ma2024understanding}. 
A particularly promising approach is \re{the} Physics-Informed Neural Networks (PINNs), which embed physical laws—such as differential equations—directly into the loss function. This allows the learning process to be constrained according to well-established physical dynamics~\cite{raissi2019physics}. Recent advances have extended PINNs to brain dynamics, demonstrating improved parameter inference and enhanced interpretability of neural processes~\cite{sotero2024parameter}. 
Additionally, the physics-structure-preserving neural networks, such as the Hamiltonian Neural Network (HNN) and Lagrangian Neural Network (LNN), \re{have begun} to show promise in improving long-term prediction capabilities~\cite{greydanus2019hamiltonian, cranmer2020lagrangian}. While the direct application of these physics-based models to the creation of surrogate brains presents challenges, due to the complexity, adaptability, and non-conservative nature of the brain, \re{this framework provides} a systematic approach for examining neural nonlinearity, energy optimization, and long-term stability~\cite{cranmer2020lagrangian, greydanus2019hamiltonian, kim2018recognition}.

In summary, gray-box models offer a compelling framework for integrating both mechanistic insights \re{and} data-driven learning, thereby facilitating more precise simulations of brain-like behavior. By integrating neuroscientific and physical priors, these models establish a robust foundation for modeling neural dynamics and constructing surrogate brains. Nevertheless, the integration of such priors is accompanied by challenges. The trade-off between model interpretability and expressiveness continues to be a persistent issue, as does the lack of strong theoretical underpinnings in certain instances. Despite these challenges, ongoing advancements in inverse problem-solving, parameter estimation, and prior engineering are expected to unlock the full potential of gray-box models, thereby enhancing our comprehension of neural dynamics and advancing the field of computational neuroscience.

% Solving Inverse Problems of Surrogate Brain 
\section{SOLVING INVERSE PROBLEMS FOR PREDICTING BRAIN DYNAMICS}\label{sec3}
This section focuses on a systematic interpretation of the framework for solving inverse problems in brain systems. First, we investigate two mainstream paradigms for solving inverse problems—the probabilistic framework and the deterministic framework. Subsequently, from the perspective of neuroscience, we discuss the well-posedness of the solutions to inverse problems, with a particular emphasis on analyzing the existence, uniqueness, and robustness of the solutions. Finally, combined with practical application scenarios, targeted strategies for alleviating ill-posed problems are proposed.

\subsection{Solving Data-Driven Inverse Problems}
To effectively solve inverse problems that infer unknown systems from observed data, researchers have established two main frameworks: the probabilistic framework based on the Bayesian theorem and the deterministic framework based on functional analysis. 

The probabilistic framework treats the parameters to be solved as random variables~\cite{hashemi2020bayesian}, with Bayesian inference as its core \re{principle}. This framework quantifies the relationship between parameters and empirical data through the likelihood function. By integrating prior information and applying Bayes' theorem, Bayesian inference derives the parameter posterior distribution, and then a point estimate is obtained via methods like Maximum A Posteriori (MAP) estimation, \re{thereby} obtaining the model solution and quantifying its uncertainty. In practical implementation, since posterior distributions often lack an analytical form, their computation relies on sampling methods such as Markov Chain Monte Carlo (MCMC) algorithms to approximate the posterior distribution~\cite{andrieu2003introduction}. However, when dealing with complex models (\eg{personalized brain models}), MCMC incurs high computational costs. In this context, probabilistic machine learning methods such as neural density estimators~\cite{papamakarios2017masked} and variational inference~\cite{graves2011practical} have emerged as more efficient alternatives. It is worth noting that the free energy principle~\cite{friston2010free} posits that organisms optimize recognition density by minimizing free energy, thereby approximating the posterior distribution. This mechanism is highly consistent with the core logic of variational inference, providing a theoretically sound and biologically meaningful framework for applying variational inference to posterior inference tasks in complex biological systems such as personalized brain models.

The core advantages of Bayesian inference lie in the fact that the prior distribution can incorporate the knowledge of domain experts, the likelihood function can accurately characterize the data generation mechanism and restore the real process, and the posterior distribution can quantify the uncertainty of the solution. These advantages naturally align with white-box models. However, in black-box models, the parameters lack interpretable mechanistic meanings, which forces the prior to default \re{on} uninformative assumptions. Due to the absence of an explicit generative structure, the likelihood relies on sampling-based approximations, the posterior also fails to anchor meaningful biological or physical interpretations. Therefore, Bayesian inference is more widely applied in solving white-box models\cite{jha2022fully, hashemi2020bayesian, friston2010free}.

In the deterministic framework, the parameters to be solved are regarded as definite entities. The deviation between the output of the forward model and the observed data is quantified by defining the data fidelity term $d(F_{\theta}(\mathbf{x}),\dot{\mathbf{x}})$, where $d$ is a distance function, its form depends on the nature of the noise term~\cite{hansen2010discrete}. Under the assumption of Gaussian noise, the least squares norm is usually employed as the distance metric. To incorporate prior knowledge, a regularization term $R(\theta)$ needs to be introduced (which will be elaborated in the following subsection). Subsequently, the data fidelity term and the regularization term are summed with weights to construct the objective function:
\begin{equation}\label{ObjectiveFunc}
\hat{\theta} \in \arg\min_{\theta} \left( d(F_\theta(\mathbf{x}), \dot{\mathbf{x}}) + \lambda R(\theta) \right).
\end{equation}
Here, $\lambda$ is a hyperparameter used to balance the trade-off between data fitting and the strength of prior constraints. Its value is often selected based on empirical knowledge and can also be obtained through cross-validation or the L-curve method~\cite{hansen1999curve}. Finally, the parameter solution is obtained by minimizing this objective function. In terms of computational implementation, the deterministic framework uses numerical optimization algorithms to solve  the minimum of the objective function. The choice of optimization algorithm needs to be flexibly adjusted according to the nature of the loss function and the form of the problem. Table~\ref{tab:algo-summary} summarizes common scenarios and the corresponding selection of optimization algorithms. For a more detailed analysis, please refer to the supplementary materials.

The deterministic framework lacks \re{the} quantification  of solution uncertainty, thus potentially posing certain risks in scientific inference and clinical applications. However, it does not require sampling from distributions and instead conducts calculations directly through optimization methods, offering high computational efficiency. Moreover, it adapts well to both black-box and gray-box models. These advantages lead to its widespread application in the field of neuroscience \cite{luo2025mapping, valente2022extracting, song2016training}. 

In recent years, the integration of the two frameworks has gradually emerged. For example, in the work of Sip et al. ~\cite{sip2023characterization}. \re{First,} variational inference is used to construct the approximate posteriors of system states, regional parameters, and other latent variables  from observational and structural data. Then, the expectations and variances of these latent variables are used as inputs to a neural network for learning dynamical systems. Finally, by constructing a minimized evidence lower bound, the network parameters of the dynamical system function and the distribution parameters of the latent variables are jointly optimized. Through this combination, the probabilistic framework is utilized to capture the uncertainty of latent variables, while the efficient fitting ability of deterministic neural networks is utilized to characterize complex dynamic rules. Ultimately, this approach provides solutions for inverse problems that possess both probabilistic characteristics and mechanical interpretability. This is also one of the implementation methods of the gray-box model mentioned earlier.

\begin{table*}[t]
\centering
\caption{Optimization algorithms}
\label{tab:algo-summary}
\resizebox{\textwidth}{!}{\renewcommand{\arraystretch}{1.3}\setlength{\extrarowheight}{2pt}%
\begin{tabular}{|l|ll|l|}
\hline
Application Scenario &
  \multicolumn{2}{l|}{Core Algorithm} &
  Update Formula / Key Mechanism \\ \hline
\multirow{3}{*}{Unconstrained} &
  \multicolumn{2}{l|}{Gradient Descent (GD)} &
  $\mathbf{x_{k+1}} = \mathbf{x_k} - \alpha \nabla J(\mathbf{x_k})$ \\ \cline{2--4} 
 &
  \multicolumn{2}{l|}{Newton's Method} &
  $\mathbf{x_{k+1}} = \mathbf{x_k} - H^{-1} \nabla J(\mathbf{x_k})$ \\ \cline{2--4} 
 &
  \multicolumn{2}{l|}{Levenberg-Marquardt Algorithm} &
  $\mathbf{x_{k+1}} = \mathbf{x_k} - (H + \lambda I)^{-1} \nabla J(\mathbf{x_k})$ \\ \hline
\multirow{3}{*}{Constrained} &
  \multicolumn{2}{l|}{Penalty Function Method} &
  $J(\mathbf{x}) + \sigma \sum_i P(C_i(\mathbf{x}))$ \\ \cline{2--4} 
 &
  \multicolumn{2}{l|}{Interior Point Method} &
  $J(\mathbf{x}) - \mu \sum_i \ln(-C_i(\mathbf{x}))$ \\ \cline{2-4} 
 &
  \multicolumn{2}{l|}{Sequential Quadratic Programming} &
  Transformed into quadratic programming subproblems \\ \hline
\multirow{3}{*}{High-Dimensional} &
  \multicolumn{2}{l|}{BFGS} &
  $H_{k+1} = \left(I - \rho_k s_k y_k^T\right) H_k \left(I - \rho_k y_k s_k^T\right)+ \rho_k s_k s_k^T$ \\ \cline{2--4} 
 &
  \multicolumn{2}{l|}{L-BFGS} &
  \begin{tabular}[c]{@{}l@{}}Based on the most recent m pairs of $(s_k, y_k)$ to \\ recursively approximate the inverse of the Hessian\end{tabular} \\ \cline{2--4} 
 &
  \multicolumn{2}{l|}{Conjugate gradient method} &
  $\mathbf{d_{k+1}}=-\nabla J(\mathbf{x_{k+1}})+\beta_k \mathbf{d_k}$ \\ \hline
\multirow{2}{*}{Large-Scale Data} &
  \multicolumn{2}{l|}{Stochastic Gradient Descent (SGD)} &
  Mini-batch samples replace full dataset \\ \cline{2--4} 
 &
  \multicolumn{2}{l|}{Adam Optimizer} &
  Momentum smoothing and adaptive step size \\ \hline
\multirow{2}{*}{\begin{tabular}[c]{@{}l@{}}Non-Differentiable \\ Difficult to Compute\end{tabular}} &
  \multicolumn{2}{l|}{Nelder-Mead Simplex Method} &
  \begin{tabular}[c]{@{}l@{}}Updated through reflection/expansion/contraction \\ operations of the simplex\end{tabular} \\ \cline{2--4} 
 &
  \multicolumn{2}{l|}{Surrogate Gradient Method} &
  \begin{tabular}[c]{@{}l@{}}Smooth surrogate functions approximate \\ non-differentiable functions\end{tabular} \\ \hline
 &
  \multicolumn{2}{l|}{Trust Region Framework} &
  $\rho_k = \frac{J(\mathbf{x_k}) - J(\mathbf{x_k} + \mathbf{d_k})}{\mathbf{{m_k(0)} - \mathbf{m_k(\mathbf{d_k})}}}$ \\ \hline
\end{tabular}
}
\end{table*}

\subsection{Well-posedness of the learning problem}

\begin{figure}[htbp]
    \centering
    \includegraphics[width=\textwidth]{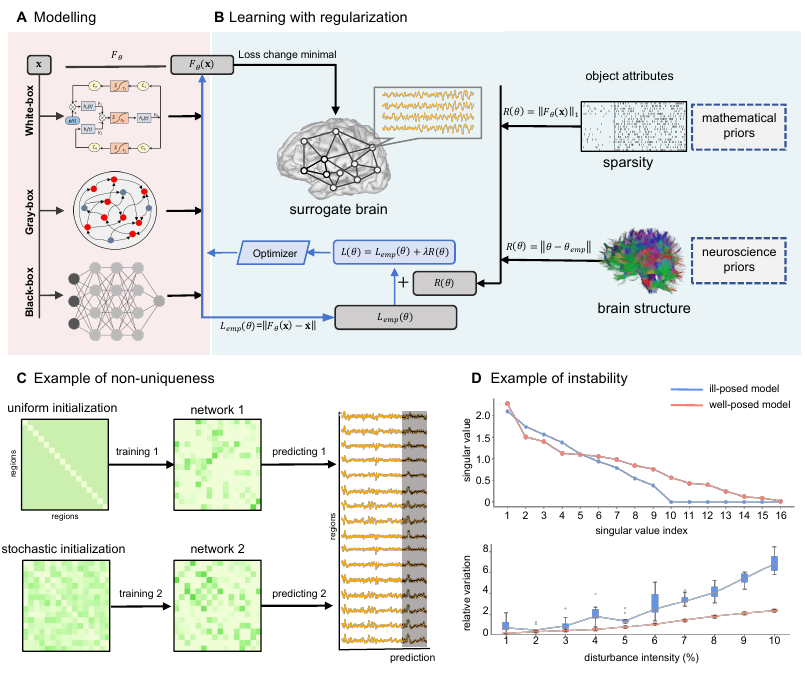}
    \caption{Solving inverse problem.  
    \textbf{A. Modelling.} 
    \textbf{B. Learning with regularization.}  Priors on object attributes are added as regularization to the loss function, and the optimizer solves the learning problem. 
    \textbf{C. Example of non-uniqueness.}  Different initializations in the optimization algorithm yield distinct solutions with comparable model performance.
    \textbf{D. Example of instability.}  Stability analysis via singular value spectra (top) and perturbation testing (bottom). 
    }
    \label{fig:procedure}
\end{figure}

A core challenge in solving the brain's inverse problem stems from its intrinsic nature: given the brain's immense number of interacting components, it is fundamentally a high-dimensional and nonlinear optimization problem (Fig.~\ref{fig:procedure} \textbf{A,B}). The objective function of such a problem is prone to having multiple local minima. Thus, during iterative optimization, different initial values may lead to distinct solutions that all fit the observed data well ( Fig.~\ref{fig:procedure} \textbf{C}). This multiplicity of solutions directly results in ambiguity in the interpretation of neural mechanisms, making model-based physiological inferences unreliable. Additionally, neural data observed in the real world inevitably contain noise. If minor observational noise can cause drastic fluctuations in the solutions, it will limit the reliability of surrogate brain models in practical scenarios such as clinical diagnosis and neuromodulation. To ensure the credibility of the solutions, we introduce well-posedness as a prerequisite that modeling should satisfy. Well-posedness originates from Hadamard's classical characterization of mathematical problems. Here, we provide an accessible introduction from a neuroscience perspective (detailed mathematical definitions can be found in the Supplementary Material).

Mathematically speaking, three key criteria are usually \re{considered} in the inverse problem for AI surrogate brain. \textbf{Existence}: \re{given} observational data, there exists at least one set of latent variables such that the predicted data generated by the surrogate brain through these variables matches the observational data within acceptable error margins, and these latent variables must be biologically plausible. \textbf{Uniqueness}: For a given system and data, only one surrogate brain model is obtained. \textbf{Stability}: When observational data is subject to natural physiological perturbations, the model solution undergoes only minor changes within acceptable limits.

General well-posedness criteria stem from the properties of the objective function \eqref{ObjectiveFunc} (denoted as \(J(\theta)\). If \(J(\theta)\) is coercive, its minimizing sequence must be bounded. In compact spaces, bounded sequences contain convergent subsequences, and when the objective function is lower semi-continuous, the limit point of such subsequences is provably a global optimal solution. Additionally, if the loss function is strictly convex, the global optimal solution obtained above is unique. Regarding solution stability, the objective function can be extended to a bivariate function \(J(\theta, \mathbf{y})\), where \(\theta\) is the solution and \(\mathbf{y}\) is the observational data. If \(J\) is Lipschitz continuous with respect to data \(\mathbf{y}\) and strongly convex with respect to solution \(\theta\), then it can be proven that the solution \(\theta(\mathbf{y})\) is Lipschitz continuous with respect to \(\mathbf{y}\) — meaning solution fluctuations are controlled by perturbations in observational data. In probabilistic settings, the Lipschitz condition metric should be revised to the Hellinger distance\cite{dashti2011uncertainty,fang2023reservoir}.

In different modeling scenarios, the well-posedness theory takes on more specific forms. When inverse problems can be modeled by linear operators as \(K\mathbf{x} = \mathbf{y}\) , the theory of their well-posedness can be found in any functional analysis textbook. Here, we introduce only the condition number as an index  to measure a linear model's resistance to perturbations. For an invertible linear operator \(K\), the condition number is defined as \( \operatorname{cond}(K) = \|K\|\|K^{-1}\| \) in the operator norm sense (equivalent to the ratio of maximum to minimum singular values under the 2-norm). If \( \operatorname{cond}(K) \) is small (\eg{close to 1}), minor input perturbations induce only slight solution fluctuations, indicating stability. When inverse problems are modeled as \re{the} biophysical equation \eqref{Jirsa}, relevant progress has also been made. Lang et al.(2024)\cite{lang2024interacting} used a framework of joint graph-kernel optimization based on alternating least squares to solve such problems, and proved that solutions are identifiable when data satisfies rank joint coercivity conditions. This condition further ensures that the matrices used in numerical computations were well-conditioned, with their minimum singular values bounded below by a positive constant(condition number \re{is} not excessively large). 

However, in high-dimensional nonlinear scenarios with noisy data, well-posedness theory may be difficult to satisfy, and AI models, even more so, lack such \re{theories of well-posedness}. In these circumstances, numerical verification methods, along with theoretical analysis, serve as an important approach. By experimentally observing the actual behavior of the model, the characteristics of well-posedness can be indirectly evaluated. Among them, multi-initial-value inversion sets multiple initial conditions to simulate the evolution paths of the system starting from various points, and judges whether the model has reliable solutions by comparing the convergence and consistency of the final results (Fig.~\ref{fig:procedure} \textbf{C}); noise perturbation testing artificially inputs random noise into the system to simulate the inevitable interference factors in reality, and observes the robustness of the model under perturbations(Fig.~\ref{fig:procedure} \textbf{D}). If the output results still fluctuate within an acceptable range, it indicates that the model has good stability. Sensitivity analysis systematically varies specific key elements, such as critical input features, core model parameters, or data distribution attributes, to precisely gauge how outputs respond. It highlights cases \re{in which} minor perturbations to these targeted elements trigger significant output fluctuations, pinpointing vulnerabilities that may compromise stability.

% 现在
It is important to note that the parameter solutions in white-box models and the inferred latent variables in \re{gray}-box models typically incorporate priors and often possess biological interpretability. Interpreting these parameters directly supports the model’s validity. Therefore, the requirements for \re{the} uniqueness and stability are natural. However, in black-box models, the weight parameters of neural networks often lack practical biological meaning; overparameterized networks may also lead to multiple sets of equivalent solutions (i.e., different parameters yielding consistent prediction performance). Therefore, the general requirements for uniqueness and stability of physical parameters, as outlined in inverse problem well-posedness theory, are not applicable to overparameterized AI models. For black-box models, the focus of evaluation should shift to aspects such as prediction performance, robustness to noise, and generalization ability. Moreover, when black-box models are used as surrogate brains, we still expect the latent representations of their outputs to be stable, indicating that the model has found an appropriate embedding space for representing neural dynamics data.

% \begin{figure*}[htbp]
%   \noindent
%   \hspace*{-0.13\paperwidth}% Shift the entire minipage left
%   \begin{minipage}{\dimexpr0.78\paperwidth}% Minipage spans the full page width
%     \includegraphics[width=0.78\paperwidth]{Figures/Figure_R1/R1_figure3.pdf}%
%     \captionsetup{justification=raggedright, singlelinecheck=false}
%     \caption{Solving inverse problem.  
%     \textbf{A. Modelling.} 
%     \textbf{B. Learning with regularization.}  Priors on object attributes are added as regularization to the loss function, and the optimizer solves the learning problem. 
%     \textbf{C. Example of non-uniqueness.}  Different initializations in the optimization algorithm yield distinct solutions with comparable model performance.
%     \textbf{D. Example of instability.}  Stability analysis via singular value spectra (top) and perturbation testing (bottom). 
%     }
%      \label{fig:procedure}
%     \end{minipage}
%     \end{figure*}

\subsection{Mitigation of Ill-posed Problems}

For most systems, well-posedness theory provides sufficient, but not necessary conditions. Due to unique challenges in neuroscience research—strong nonlinearity of neural dynamics, multi-scale coupled physical properties, and inherent noise and uncertainty in observational data—these well-posedness conditions may be difficult to \re{meet}. At the same time, the results of numerical verification are more likely to indicate that the model is ill-posed. Given the pervasiveness of this ill-posed phenomenon, we now introduce some strategies for addressing ill-posed problems.

Regularization theory is precisely the key approach to solving such problems. Its core idea is to introduce prior constraints by adding a regularization term, presetting the structural features of parameter solutions or model behavior to guide the optimization process. This ensures the stability of solutions when data is limited or disturbed by noise. \re{In parallel,} the constraints narrow the solution space, making the solutions conform to physiological characteristics. Even if the solutions are still not unique, they can be considered approximately unique in the equivalent sense of neuroscience. It primarily functions through two approaches: one \re{directly constrains} the model parameters; the other \re{constrains} the model output, indirectly constraining the parameters through the correlation between the output and the parameters. Here we first introduce two types of explicit regularization terms, which are derived from mathematical priors and neuroscience knowledge priors, and then briefly mention implicit iterative regularization and alternative strategies for mitigating ill-posed problems of AI surrogates.

Mathematical prior regularization uses mathematical structures to constrain the expected properties of the solution or the model's behavior. Constraints on white-box model parameters and gray-box model latent variables often depend on their biological properties, which we will explain in detail in the section on biological prior regularization. Here, we introduce \re{the concept of} constraining the model's output and constraining the weights of the black-box model. Common forms can be divided into the following categories.

Sparsity Constraints \re{encoding} the regularization term as the L1 norm, enabling the model to retain only key components for computation~\cite{yu2022learning,tibshirani1996regression}. In neural networks, sparsity constraints applied to parameters can eliminate redundant connections caused by noise~\cite{han2015deep}, while applying constraints at the output level ensures that model prediction results align with the sparse characteristics of neural activities. Energy Constraints \re{involve} using the L2 norm as the regularization term, which essentially restricts the sum of the squared magnitudes of parameters, keeping the fluctuations of parameter magnitudes within a reasonable range. This is \re{one of} the most commonly used regularization method in neural networks. At the parameter level, energy constraints can prevent parameters from growing \re{without restriction}, avoiding the model from overfitting the noise in the data; at the output level, they ensure that model outputs conform to the physiological thresholds of neural responses. Additionally, the introduction of the L2 norm can enhance the convexity of the objective function, facilitating the acquisition of a unique optimal solution\cite{goodfellow2016deep}. Regularization terms for global smoothness constraints primarily act at the output level and are often defined as the L2 norm of the derivative of the model output with respect to the input~\cite{belkin2006manifold}. Their core function is to penalize drastic changes in the output with respect to the input, forcing the model to learn a smooth mapping between inputs and outputs, thereby enhancing the model's robustness to noisy neural data. The regularization term of Total Variation (TV) is typically defined as the L1 norm of the gradient operator, which suppresses local high-frequency fluctuations while preserving significant edge features. This property \re{aligns well with} the characteristic in neuroscience that neural data is locally continuous and has distinct regional boundaries\cite{dohmatob2014benchmarking}. At the parameter level, TV regularization can constrain the local changes in the weight matrix, preventing irregular jumps in neural connection patterns; at the output level, it can retain important edge information when decoding brain activities, ensuring consistency with the boundaries of known brain functional regions.

% 结构，功能，第一性原理。多cite几篇吧
Neuroscience knowledge-based prior regularization systematically encodes prior knowledge from neuroscience—including anatomical structures, physiological mechanisms, and neural dynamics—into regularization terms. This ensures models not only yield numerical predictions but also conform to biological principles. Here, we elaborate on constraint methods applied to the parameter solutions of white-box models, the latent variables of gray-box models (collectively referred to as ``solutions'' hereafter), and the outputs of all models.

First, it should be noted that when these solutions exhibit corresponding characteristics, the aforementioned mathematical prior regularization can be directly applied to impose constraints such as sparsity and energy constraints. Additionally, based on structural and functional knowledge, prior information about the solutions can be embedded \re{in} the loss function. At the anatomical level, integrating structural priors derived from techniques like DTI minimizes deviations between the model and structural connectivity, ensuring that latent representations align with the brain’s physical (anatomical) structure~\cite{chen2017features,achterberg2023spatially}. Meanwhile, region-specific constraints (\eg{penalizing violations of known functional hierarchies or functional connectivity patterns}) can further guide the model to comply with cortical and subcortical structural organizations~\cite{kong2021sensory}. Functional priors (such as task-specific modularity) can enhance the model's interpretability and performance~\cite{guo521519}. Modularity-based penalty terms prompt the model to adjust its internal representations. \re{These adjustments reflect distributed and specialized activity patterns in the brain. This aligns with the characteristics of functional segregation and integration observed in neural systems~\cite{yao2024brain}.} For model outputs, the primary approach is to adopt dynamic equations from first-principles physical models as constraints, incorporating \re{the} residuals of these physical models into the objective function. This forces the model to follow physical laws during training, effectively improving the model’s plausibility and predictive accuracy~\cite{sotero2024parameter}.

The essence of mitigating the ill-posedness of inverse problems lies in constraining the solution space, which is not limited to being achieved through explicit regularization terms \re{alone}. Iterative regularization precisely stops the iteration before the model overfits the noise, simultaneously capturing the optimal solution and achieving regularization~\cite{kaltenbacher2008iterative} . Since it \re{avoids the need for} constructing and storing high-dimensional regularization terms, it is suitable for large-scale problems. The Direct Sampling Method explores the parameter space through random sampling or geometric partitioning, reducing the dependence on local noise \re{and} stabilize the solution output. Moreover, it does not rely on gradients, making it \re{especially} suitable for ill-posed problems with strongly nonlinear and non-smooth solutions~\cite{ning2025direct}.
Data prior guidance employs techniques such as parameter freezing and feature distillation to introduce the latent patterns in large-scale data as data priors into the model. \re{This approach constrains} the parameter distribution and output features, and reduces the sensitivity to noise and sparse data. Dynamic adjustment strategies~\cite{lecun2015deep} in neural network training use Dropout for random sparsification to reduce overfitting, batch normalization to stabilize the input distribution, and data augmentation to expand data diversity. \re{Together, these strategies guide} the model to learn essential laws, \re{ultimately contributing to} the mitigation of ill-posedness.

% Finally, we point out that the above strategies are only mitigations of ill-posed problems and cannot automatically guarantee the well-posedness of the inverse problem. However, take regularization  as a key example，extensive engineering practice has empirically validated that when regularization strategies align deeply with problem characteristics and hyper-parameters are reasonably selected, regularization significantly reduces ill-posedness. In practical applications, regularized solutions not only meet accuracy requirements but also demonstrate consistency with experimental observations and physical laws, providing reliable foundations for engineering decisions. Thus, despite theoretical limitations, regularization remains the core methodology for mitigating ill-posedness and obtaining practical solutions in inverse problems.

Finally, we emphasize that the aforementioned strategies serve primarily as mitigations for ill-posed problems and cannot automatically guarantee the well-posedness of the inverse problem. However, consider regularization as a key example: extensive practical experience has empirically demonstrated that when regularization strategies are deeply aligned with the characteristics of the problem and hyperparameters are appropriately chosen, they substantially alleviate ill-posedness. In practical applications, regularized solutions not only satisfy accuracy requirements but also exhibit consistency with experimental observations and physical principles, thereby offering a reliable basis for engineering decision-making. Thus, despite inherent theoretical limitations, regularization remains a cornerstone methodology for reducing ill-posedness and deriving useful solutions in inverse problems.

\section{EVALUATION}

Once the inverse problem of a neural dynamical system has been addressed, the next step is to evaluate the trained model. The goal is to determine whether the model captures the essential dynamics of neural systems and provides interpretable representations, \re{functioning} as surrogate brains. Neural dynamics are often only partially accessible. \re{As a result,} evaluation typically relies on comparing model-generated and empirical data.
Evaluation can be \re{approached} from two complementary perspectives. From a mathematical standpoint, similarity analyses across different metric spaces highlight signal fidelity and reveal distinct system characteristics. From a neuroscientific perspective, the focus shifts to functional alignment, assessing whether models reproduce task-related activity and support consistent decoding.
In what follows, we review evaluation metrics from both perspectives.

\begin{figure}[htbp]
    \centering
    \includegraphics[width=\textwidth]{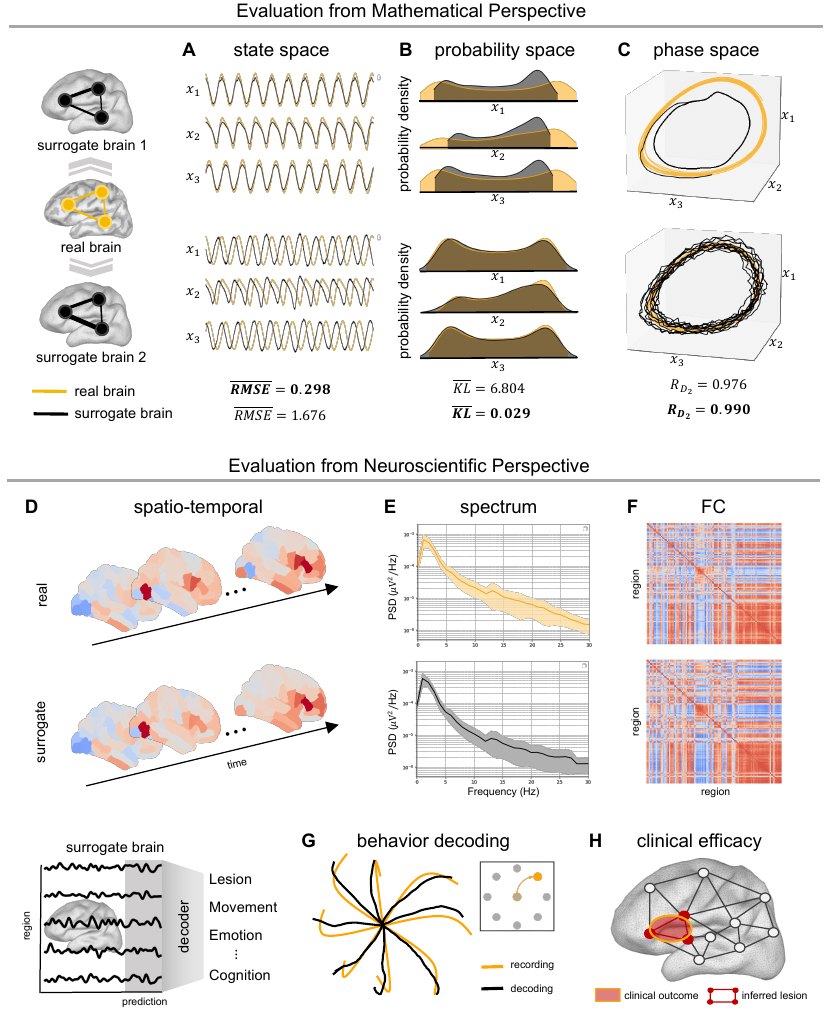}
    \caption{\textbf{Model evaluation from mathematical and neuroscientific perspectives.}
   \textbf{A–C| Mathematical perspective.}
   Synthetic data were generated with the stochastic Jansen–Rit model~\cite{Ableidinger2017ASV}. The upper and lower rows correspond to two surrogate brains derived from the same underlying system, evaluated across three representational spaces: 
   \textbf{A.} state-space trajectories(mean root-mean-square error, $\overline{\text{RMSE}}$); \textbf{B.}probability distributions(mean Kullback–Leibler divergence, $\overline{\text{KL}}$), \textbf{C.} phase-space dynamics(relative correlation dimension, $R_{D_2}$).
   Surrogate 1 achieves lower \(\overline{\text{RMSE}}\), while Surrogate 2 performs better in probabilistic and topological metrics.
   \textbf{D–H| Neuroscientific perspective.} Metrics include \textbf{D.} spatial–temporal similarity, \textbf{E.} spectral similarity, \textbf{F.} FC similarity, \textbf{G.} behavioral decoding accuracy, and \textbf{H.} clinical efficacy in predicting lesion outcomes.
   Panels D and E focus on the surrogate brain's ability to replicate biological signal characteristics, Panels G and H evaluate its capability to support downstream tasks.
   Yellow traces denote real brain data; black traces represent surrogate brain outputs.
    }
    \label{fig:evaluation}
\end{figure}

\subsection{Mathematical perspective}

The mathematical evaluation of the inverse problem \re{in} neural dynamical systems often begins with assessing data similarity across various metrics, as this \re{reveals} distinctive features of the investigated system~\cite{edgar2008measure}. A standard approach is to compute pointwise errors such as Mean Squared Error (MSE)~\cite{ding2024deep}, a simple and widely used measure of temporal similarity in machine learning (Fig.\ref{fig:evaluation} \textbf{A}).

However, their simplicity is a significant limitation: they fail to capture higher-order geometric and topological properties of complex systems. For instance, even if a generated system replicates the topological structure of the original trajectory, minor variations in initial conditions or phase shifts can lead to large norm errors that do not reflect meaningful dynamical differences (Fig.~\ref{fig:evaluation} \textbf{A-C}). This issue is particularly relevant for the brain, a high-dimensional and highly sensitive dynamical system where small perturbations can drastically alter dynamics~\cite{breakspear2017dynamic}. Therefore, while a low MSE may suggest good temporal alignment, it does not necessarily guarantee an accurate representation of the brain's underlying structural or dynamical properties~\cite{durstewitz2023reconstructing}. This limitation highlights the need for alternative evaluation strategies that better capture the complexity of brain dynamics. Beyond pointwise error metrics, goodness-of-fit measures are another important approach for model evaluation. Metrics such as Explained Variance (EV) and the Coefficient of Determination ($R^2$) quantify how much of the data variance \re{the model accounts for}, without \re{relying} on specific features~\cite{volkmann2024scalable, jiang2024identifying}. These measures complement pointwise errors by offering a broader perspective on model accuracy and providing a more intuitive view of how well the generated dynamics reproduce empirical observations.

% Beyond pointwise error metrics, probabilistic approaches offer a complementary view by analyzing the global properties of data distributions(Fig.~\ref{fig:evaluation} \textbf{B}). 
In addition to state space evaluations, probabilistic approaches provide complementary insights by analyzing the global properties of data distributions (Fig.\ref{fig:evaluation} \textbf{B}).
These methods assess the overlap between real and surrogate brains by comparing probability distributions, providing insights into the structural characteristics of the system's dynamics. \re{Commonly} used probabilistic metrics include Kullback-Leibler divergence, Wasserstein distance, and Hellinger distance~\cite{koppe2019identifying, karpowicz2025stabilizing, gilpin2020deep, zhang2024sparse}. 
For instance, the Wasserstein distance not only measures positional discrepancies between distributions but also captures their geometric shapes, making it a powerful metric for evaluating high-dimensional data ~\cite{chung2023unified}. 
Such probabilistic measures are especially important in neuroscience, where stochastic variability and noise can profoundly influence neural trajectories, necessitating robust methods to evaluate model fidelity.

Topological and geometric properties are fundamental to characterizing the long-term behavior of dynamical systems. Key features, such as chaos, attractors, and limit cycles govern the system's evolution and dynamic responses~~\cite{durstewitz2023reconstructing}. The brain is often hypothesized to operate at the edge of chaos, a critical state where order and adaptability are balanced. This state is thought to optimize signal propagation and enhance the system’s ability to respond flexibly to a dynamic environment~\cite{o2022critical}.
A metric for quantifying sensitivity to initial conditions is the maximal Lyapunov exponent (MLE). \re{This metric indicates} whether a generated system accurately replicates the brain's critical dynamical features, including its hypothesized operation near the edge of chaos~\cite{morales2021optimal}.
Complementing this, the fractal dimension measures the geometric complexity and self-similarity of attractors, providing insights into the structure and dynamics of brain activity~\cite{breakspear2017dynamic, stylianou2022scale}. Comparable fractal dimensions between real and surrogate systems suggest that the surrogate not only preserves system complexity but also reproduces its topological organization (Fig.~\ref{fig:evaluation} \textbf{C}).
In addition to these approaches, persistent homology, as a tool in topological data analysis, offers a powerful method for studying the multi-scale structure of state spaces by capturing the birth and death of topological features such as connected components, loops, and voids across different scales~\cite{ turkes2022effectiveness}. This approach is particularly suited for analyzing complex, high-dimensional brain networks, as it extracts robust topological features across scales and is less sensitive to noise~\cite{liang2017structure}.

\subsection{Neuroscientific perspective}

From a neuroscientific perspective, a central question is whether surrogate brains preserve the physiological characteristics of real neural systems. Evaluation typically considers spectrotemporal properties, functional connectivity (FC), and task-related performance. These evaluations provide insights into fundamental features, dynamic properties, and task-specific performance, \re{all} reflecting the functional and structural characteristics of neural systems (Fig.~\ref{fig:evaluation} \textbf{D-H})~\cite{kurtin2023moving}. To address these concerns effectively, evaluation frameworks must extend beyond traditional mathematical error analyses by incorporating neuroscience-related metrics.

Unlike purely mathematical evaluations, neuroscientific assessments emphasize the consistency of spatial-temporal activation patterns, which capture the model’s ability to reproduce functional activity across both specific brain regions and the whole brain (Fig.\ref{fig:evaluation} \textbf{D})\cite{mi2021connectome, lin2022imaging}. High similarity in these patterns indicates that the surrogate brain retains the functional coordination characteristic of neural systems.
Dynamic properties are further examined through spectral and FC analyses. Spectral analysis characterizes rhythmic activity across scales~\cite{capilla2022natural, rodrigues2021hnpe}, \re{ranging} from micro-level neuronal firing to macro-level oscillatory states associated with EEG frequency bands (Fig.\ref{fig:evaluation} \textbf{E}). In parallel, FC analysis evaluates \re{neural activity coordination} across regions by quantifying spatiotemporal dependencies. Static FC, based on averaged correlations, provides a global assessment of whole-brain performance(Fig.\ref{fig:evaluation} \textbf{F}), whereas dynamic FC captures time-varying connectivity and the complex temporal organization of neural systems~\cite{wei2022effects}. %Frequency-specific FC is also used to evaluate the surrogate model since neural oscillations at different frequencies support distinct connectivity patterns~\cite{li2022hierarchical}.

In practical applications, surrogate brains are typically validated \re{through} task-specific metrics that assess how well they replicate real brain functions~\cite{mathis2024decoding}. For example, researchers compare whether the surrogate brain can decode lesions~\cite{dou2023identification}, behaviors~\cite{pellegrino2023low}, emotions~\cite{shen2025dynamic}, or cognitive states~\cite{ye2023explainable} \re{aligning with} actual neural data.
For instance, in epilepsy localization, the overlap between clinically resected regions and the predicted epileptogenic zones derived from the surrogate brain serves as a key metric for assessing localization accuracy (Fig.~\ref{fig:evaluation} \textbf{I})~\cite{jirsa2023personalised, millan2024individualized}. \re{This approach links model predictions directly to clinical outcomes}, providing a robust measure of its practical utility. Similarly, task-related neural recordings \re{allow} researchers to evaluate model reliability through behavioral predictions. For example, decoding monkey movement trajectories from neural data and comparing them to observed behaviors offers a direct validation of the model's task-level reliability (Fig.~\ref{fig:evaluation} \textbf{J})~\cite{pellegrino2023low, schneider2023learnable}. Such task-specific validations highlight the model’s ability to bridge theoretical dynamics with practical utility.

By integrating mathematical and neuroscientific techniques, the surrogate brain can be evaluated in terms of accuracy, functionality, and practical applicability. Mathematical approaches, including error-based, probabilistic, and topological measures, characterize the structural and geometric fidelity of the models. Neuroscientific assessments, such as spatial-temporal activation patterns, spectral analysis, functional connectivity, and task-related decoding, ensure alignment with physiological \re{functions}. Together, these complementary perspectives provide a robust framework for refining surrogate brain models. \re{They improve} predictive stability during training~\cite{mikhaeil2022difficulty, park2023persistent}, and \re{enhance} decoding accuracy in applied settings~\cite{bernardo2024simulation, abbaspourazad2024dynamical}.
Nevertheless, evaluating \re{these} complex systems remains challenging. Different metrics often capture distinct aspects of neural data, sometimes leading to divergent conclusions. Moreover, neural signals vary across spatial and temporal scales, requiring evaluation strategies that are sensitive to these dynamics. Careful selection of metrics, tailored to the objectives of each task, is therefore essential for producing meaningful and reliable insights. To better address these challenges, we introduce a preliminary surrogate brain benchmark that incorporates ten complementary evaluation metrics to systematically assess model performance. Consistent with prior observations, we find that different metrics can yield markedly different evaluations of the same model (see Supplementary Figure S1).

Beyond fidelity in signal reconstruction, functional correspondence, and task-level decoding, surrogate brain models must also demonstrate robustness and generalizability. This requirement is especially important in real-world applications such as brain–computer interfaces, where data are often noisy or incomplete. Robustness can be tested by perturbing inputs, such as adding noise or masking temporal and spatial segments, to assess resilience against signal corruption and missing information~\cite{abbaspourazad2024dynamical}. Generalizability, in turn, relies on rigorous cross-validation procedures. K-fold cross-validation supports reliable performance estimation, while leave-one-subject-out approaches are particularly suitable for small-sample neuroscience datasets~\cite{kunjan2021necessity}. For models requiring hyperparameter tuning, nested cross-validation provides unbiased performance estimates~\cite{li2021neural, lumumba2024comparative}. These multi-level assessments of fidelity, robustness, and generalizability establish a comprehensive framework for benchmarking surrogate brain models.

\section{APPLICATIONS}
\begin{figure}[htbp]
    \centering
    \includegraphics[width=\textwidth]{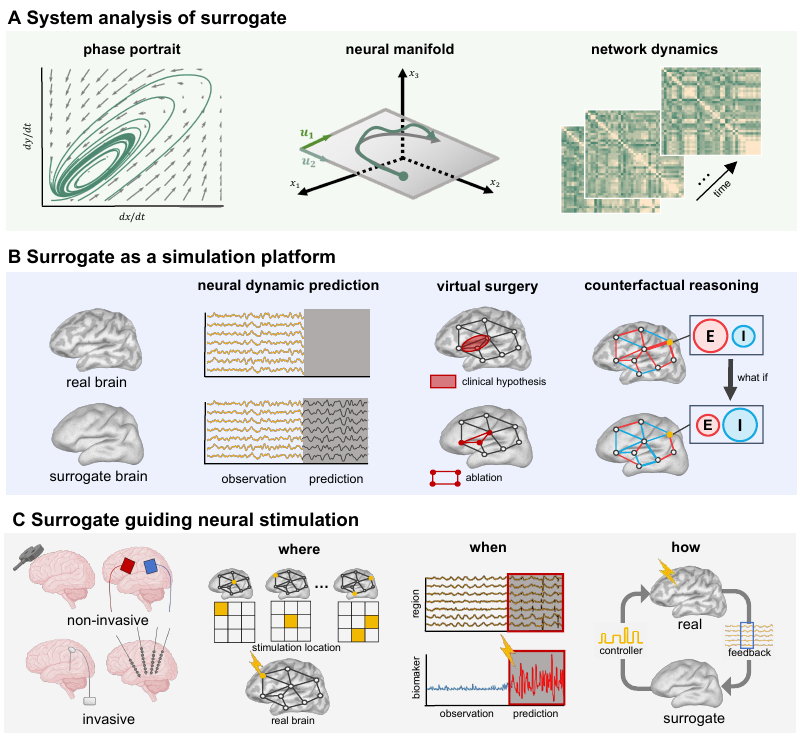}
    \caption{Applications of surrogate brain. 
    \textbf{A. System analysis of surrogate brain.}  
    The surrogate allows direct inspection of fundamental system properties, including phase portrait exploration, neural manifold geometry, and network dynamics profiling. 
    \textbf{B. Surrogate as simulation platform.}  
    As an in-silico testbed it enables neural dynamics prediction, virtual surgery (\eg{locating epileptogenic foci to aid clinical planning}), and counterfactual experiments such as perturbing connection weights or the excitation–inhibition ratio to probe their functional roles.  
    \textbf{C. Surrogate guiding neural stimulation.}  
    For both non-invasive (\eg{TMS, tES}) and invasive (\eg{DBS, SEEG}) stimulations, the surrogate brain helps determine \emph{where} to stimulate, \emph{when} to target, and \emph{how} to optimize stimulation parameters.}
    \label{fig:application}
\end{figure}

The \textbf{surrogate brain}, a neural dynamical system that generates brainlike spatiotemporal dynamics and physiological behaviors, offers broad opportunities for advancing both basic neuroscience and translational applications (Fig.~\ref{fig:application}).

% \begin{figure*}[htbp]
%   \noindent
%   \hspace*{-0.13\paperwidth}% Shift the entire minipage left
%   \begin{minipage}{\dimexpr0.78\paperwidth}% Minipage spans the full page width
%     \includegraphics[width=0.78\paperwidth]{Figures/Figure_R1/R1_figure5.pdf}%
%     \captionsetup{justification=raggedright, singlelinecheck=false}
%     \caption{Applications of surrogate brain. 
%     \textbf{A. System analysis of surrogate brain.}  
%     The surrogate allows direct inspection of fundamental system properties, including phase portrait exploration, neural manifold geometry, and network dynamics profiling. 
%     \textbf{B. Surrogate as simulation platform.}  
%     As an in-silico testbed it enables neural dynamics prediction, virtual surgery (\eg{locating epileptogenic foci to aid clinical planning}), and counterfactual experiments such as perturbing connection weights or the excitation–inhibition ratio to probe their functional roles.  
%     \textbf{C. Surrogate guiding neural stimulation.}  
%     For both non-invasive (\eg{TMS, tES}) and invasive (\eg{DBS, SEEG}) stimulations, the surrogate brain helps determine \emph{where} to stimulate, \emph{when} to target, and \emph{how} to optimize stimulation parameters.}
%     \label{fig:application}
%   \end{minipage}
% \end{figure*}

\subsection{System analysis of surrogate}
A surrogate brain can be \re{considered} a parameterized representation of neural dynamics. To examine its system properties, two complementary approaches are typically employed: direct parameter analysis and perturbation-based analysis.

In the direct approach, changes in system parameters are analyzed within phase space. A typical example is NMM. By analyzing fitted parameters that correspond to physiological indicators, such as the average membrane potential, we can uncover associated brain states~\cite{jansen1995electroencephalogram}. Constructed using differential equations, NMMs allow for in-depth exploration through dynamical and bifurcation theories~\cite{hess2023generalized, brenner2022tractable}. Complementary insight can be obtained by examining the model’s attractor landscape. Analysis \re{of} its landscape—often \re{represented} as an energy or quasi-potential surface~\cite{li2013quantifying, li2014landscape}—makes it possible to describe both limit cycles (oscillations) and multiple fixed points (multistability)—two motifs thought to jointly underpin rich brain activity. Such analyses have clarified mechanisms of resting-state maintenance~\cite{freyer2011biophysical}, distributed working memory~\cite{mejias2022mechanisms}, and motor pattern generation~\cite{zhou2024revealing}.  Notably, Ye et al. used an energy-landscape plus minimum-action-path framework to show that, even in the absence of external cues, a working memory network can reside in three co-existing attractors and demonstrated how targeted \re{landscape control} can steer transitions between them~\cite{ye2023controlling}. Fitting nonlinear dynamical systems to whole-brain scales is computationally demanding, so linear approximation models are often used. In state–space form $\mathbf{x}(t+1) = A \mathbf{x}(t) + B \mathbf{u}(t)$, the transition matrix $A$ and input matrix $B$ capture state evolution and input–output mapping. Analyzing $A$ uncovers intrinsic dynamics such as oscillation frequencies and stability, while adjusting $B$ helps explain behavioral differences~\cite{ikeda2022predicting, liang2024reverse}. Linear models have also been applied to localize epileptogenic zones by quantifying changes in network connectivity~\cite{li2021neural}.

For models lacking explicit functional forms, such as neural networks, direct parameter analysis moves towards exploring their weight matrices. However, the opaque relationship between system parameters and model outputs complicates the direct monitoring of changes in network parameters~\cite{brunton2019data}. Moreover, the high-dimensional and interdependent latent dynamics within neural networks increase the complexity of analysis and interpretation. Disentangled Representation Learning offers a promising solution to these issues. It separates latent structures based on statistical or physical principles, enabling the learning of independent factors through distinct latent variables~\cite{kim2018disentangling}. For example, Miller et al. introduced information bottlenecks in RNN to encourage sparsity and independence among state variables, thereby enhancing the transparency of cognitive models~\cite{miller2024cognitive}.

Another method for analyzing system properties is through perturbation-based approaches. This indirect method, grounded in statistical physics, involves modifying specific structures or parameters and observing the resultant changes in the dynamical system. 
For example, altering the excitatory-inhibitory (E-I) activity ratio~\cite{van1996chaos} is a common perturbation method. The E-I balance is crucial for efficient neural encoding and information processing~\cite{zhou2018synaptic, antoine2019increased}. Adjusting this ratio can change the firing rates of neural populations, thereby inducing shifts in brain states~\cite{liang2024excitation}. Additionally, structural parameters play key roles in brain dynamics. For instance, in the Kuramoto oscillator model, modifying coupling strengths can induce synchronization among neurons. This framework helps explain parameter variations driven by brain heterogeneity, including differences in neural density and signal transmission speeds across regions~\cite{li2022hierarchical, sip2023characterization}. Furthermore, tracking parameter updates in network models during learning and memory processes \re{helps understand} the mechanisms of neural plasticity, revealing how the nervous system adapts over time~\cite{Basile2023meta}.

In summary, direct parameter analysis provides detailed insights into specific parameters and their roles within the system. \re{On the other hand}, perturbation-based indirect methods reveal the system's responses to changes. Combining these two approaches offers a comprehensive understanding of the system properties of surrogate brains, thereby supporting neuroscience research and clinical applications.

\subsection{Surrogate as a simulation platform}
Traditional wet-lab experiments in neuroscience are limited by long preparation times, high costs, and ethical risks, \re{especially} in clinical contexts~\cite{jirsa2023personalised}. Moreover, counterfactual testing of ``what if'' scenarios is often technically or ethically infeasible. These issues have driven the development of simulation platforms based on surrogate brains, which allow for \re{simulating} neural dynamics at a lower cost and reduced risk, offering flexibility for research that would be difficult or even impossible to conduct directly in clinical environments.

A notable example is the Virtual Epileptic Patient (VEP), a personalized whole-brain simulation platform for epilepsy surgery planning. By reconstructing patient-specific dynamics, VEP predicts intervention outcomes and optimizes strategies, showing potential to outperform traditional experience-based planning~\cite{wang2023delineating, jirsa2023personalised}. This modeling framework has been expanded into a larger platform, The Virtual Brain, and is used for research on modeling multiple neurological diseases~\cite{monteverdi2023virtual, wang2024virtual}.

Beyond clinical use, surrogate brains also enable counterfactual reasoning in theoretical research. For instance, simulations comparing ``thalamic drive and interoceptive drive'' showed that the latter significantly improved the similarity between simulated and empirical resting-state activity~\cite{lu2024imitating}. Such controlled perturbations are nearly impossible in vivo, yet become feasible through surrogate platforms.

Using the surrogate brain as a simulation platform \re{significantly} facilitates the analysis and validation of experimental research. However, most current models rely on resting-state or task-specific neural activity data for training and fail to adequately account for the dynamic responses following interventions. This limitation poses a challenge for the broader application of surrogate brains. In particular, data-driven models, as opposed to those based on first-principles, have not encountered neural data from post-intervention or ``out-of-specific-input'' scenarios during their training. This limitation makes it difficult to determine whether the surrogate brain's responses after intervention align with real brain responses, raising concerns about the reliability of simulation platforms. A promising solution lies in combining out-of-distribution (OOD) detection with uncertainty quantification techniques. OOD detection can identify ``unfamiliar'' inputs~\cite{yang2024generalized}, while uncertainty quantification assesses the confidence in the predictions~\cite{Hashemi2021OnTI, Hashemi2023AmortizedBI, Vattikonda2021IdentifyingSS}. By combining these methods, the adaptability and reliability of the surrogate brain as a simulation platform \re{are} significantly enhanced.

\subsection{Surrogate guiding neural stimulation}

Traditional neuromodulation techniques, such as deep brain stimulation (DBS), can alleviate symptoms of disorders like Parkinson’s disease~\cite{spooner2023dbs}. However, its reliance on fixed empirical parameters limits adaptability, leading to inconsistent outcomes~\cite{martinez2022dynamical, cao2023state}. Closed-loop neurostimulation incorporating neural feedback addresses this limitation, and surrogate model–guided stimulation has recently emerged as a promising direction. It centers on three key questions: where, when, and how to stimulate (see Fig.~\ref{fig:application}C).

Where to stimulate is a critical aspect of neural stimulation, as different brain regions or neural pathways may respond differently to stimulation. Computational surrogates may help identify brain regions that are likely to be more responsive to stimulation, based on an individual's functional brain networks~\cite{li2021neural}. For example, in patients with Parkinson's disease, a surrogate model can predict the optimal deep brain stimulation site based on the individual's symptoms and fMRI \re{maximizing} the improvement of motor function~\cite{boutet2021predicting, ferrea2024machine}. Similarly, for patients with depression, data-driven surrogates identify brain areas involved in emotional regulation, precisely locating stimulation sites to enhance therapeutic effects~\cite{johnson2024deep}. Lu et al. used data-driven models to identify seizure onset zone in the epileptic brain network~\cite{lu2025ukf}, laying the foundation for targeted neuromodulation.

When to stimulate plays a decisive role in its effectiveness. For example, the effectiveness of OFC stimulation is influenced by the brain's state at the time of stimulation, with outcomes varying according to neural activity associated with specific mood states~\cite{cao2023state}. 
The surrogate model requires identifying the sustained biomarker that \re{triggers} the neural stimulator in real-time~\cite{yang2025advancements}. These biomarkers serve as indicators of optimal stimulation conditions, ensuring that the intervention is effective and timely. For example, in conditions like epilepsy, the surrogate model could monitor specific frequency patterns or low-dimensional representations that are associated with an imminent seizure. Meng et al. utilized data-driven models to predict seizure onset in a low-dimensional spatiotemporal recurrent dynamics~\cite{meng2024real}. Scangos et al. used $\gamma$ oscillations as biomarkers in closed-loop treatment of severe depression~\cite{scangos2021closed}. In DBS for Parkinson's disease, the controllers adapt the stimulation parameters according to the power or duration of $\beta$ oscillations to track ongoing state changes~\cite{bouthour2019biomarkers, wang2022adaptive}.

Effective neural stimulation requires not only precise timing and location but also consideration of stimulation strategies (\eg{intensity, frequency, and pattern}). For instance, high-frequency stimulation may be more effective for modulating excitability in certain brain regions, while low-frequency stimulation could be better suited for reducing overactivity in others.
Recently, model-guided closed-loop optimal control has emerged as a promising approach for strategy design. Since explicit dynamical equations are often unavailable, data-driven surrogate models provide a practical alternative by capturing input–output relationships and predicting future dynamics~\cite{liang2022online,yang2021modelling}. Coupled with control theory, these surrogates can derive strategies that balance therapeutic goals with safety constraints, enabling adaptive and precise neuromodulation.

In real-time neuromodulation, a key challenge is balancing predictive accuracy and computational efficiency. Nonlinear models often provide higher fidelity but are difficult to implement under strict time constraints. To address this, linearization techniques such as the Deep Koopman framework project nonlinear neural data into finite-dimensional linear spaces, enabling efficient quadratic solvers without losing predictive performance~\cite{yang2018control, liang2022online, liang2024reverse}. Additionally, dimensionality reduction methods inspired by oscillator theory (\eg{Ott-Antonsen ansatz}) have been extended to neural populations, providing reduced-order models that retain stability properties and remain practical for clinical deployment~\cite{wang2023desynchronizing}.

\section{OUTLOOK AND CHALLENGES}

Surrogate brain models provide a powerful framework for linking neural dynamics with brain function, offering transformative opportunities in diagnosis, therapy, and brain–machine interfaces. Their advancement, however, is constrained by fundamental challenges in data integration, model architecture, and inferring latent dynamics. In the following, we discuss potential solutions across data, model structure, and learning algorithms, with an emphasis on advancing surrogate brain models toward accuracy, interpretability, and clinical utility.

\subsection{Multi-scale neural data}
Neural data span multiple levels, from cellular recordings to large-scale brain imaging, covering diverse spatial and temporal scales. For instance, microscale spike trains and macroscale fMRI signals differ in resolution, noise statistics, and underlying \re{biophysical} mechanisms. Directly integrating such heterogeneous data poses risks of information loss or conflicts between modalities~\cite{tang2024modal,feng2023learning}. Thus, designing systematic strategies for aligning and integrating multi-scale data remains a central challenge.
In white-box models, Lu et al. enabled cross-scale integration by combining neural priors with data assimilation to infer neuron-level parameters from macro-level measurements (EEG and fMRI)~\cite{lu2024simulation}. While biologically grounded, these models are computationally intensive and often severely ill-posed. Data-driven surrogate brain models face related but distinct challenges.
Methods such as multi-layer networks~\cite{blanken2021connecting}, shared latent spaces~\cite{tang2023explainable}, and domain adaptation~\cite{wang2022inference, li2023novel} have been proposed to capture cross-scale relationships. \re{However}, preserving interpretability and biological plausibility across levels remains unresolved. Future progress may require embedding biologically informed constraints and adopting causal representation learning to align scales without sacrificing mechanistic fidelity.

Another major challenge arises from the intrinsic variability of neural data, stemming from individual differences, experimental conditions, and dynamic task contexts, which complicates robust cross-scale integration. A promising direction is the development of personalized surrogate brain models trained from sparse individual data, initialized using population priors and adapted via sample-efficient fine-tuning (\eg{transfer learning or meta-learning}) and generative modeling for data augmentation. Such personalized approaches should be paired with explicit uncertainty quantification and out-of-distribution detection to ensure reliable deployment.

\subsection{AI model structure}

Designing an appropriate model structure is a critical challenge in constructing surrogate brains. As the bridge between raw data and functional derivatives, the model needs to capture neural dynamics and provide a framework for solving inverse problems. This influences both the accuracy of data reconstruction and the ability to infer brain dynamics from observation.

Given the brain’s complexity, the surrogate model should possess sufficient ~\textbf{representation capacity} to handle high-dimensional, large-scale data. Recent foundation models trained on large datasets have shown promise in simulating complex neural patterns and approaching human cognitive performance~\cite{carobrainlm}. However, the challenge remains to learn representative features from limited, noisy data.
\textbf{Interpretability} is another critical aspect. While deep learning models offer high accuracy, their black-box nature complicates understanding how predictions emerge. Enhancing interpretability by incorporating biologically plausible mechanisms, such as synaptic connections and feedback loops, can improve alignment with actual brain function~\cite{zhang2022excitatory}. A key open problem is how to balance \textbf{representational capacity} and \textbf{interpretability}: overly expressive models risk becoming biologically opaque, whereas overly constrained models may fail to capture the brain’s complexity. Hybrid architectures combining biologically grounded priors with data-driven flexibility, or modular architectures informed by causal representation learning, may offer a promising path forward.

Beyond mechanistic interpretability, \textbf{explainability} addresses whether model predictions and decisions are understandable and actionable \re{to} human users. For instance, when a surrogate brain recommends a treatment plan, clinicians need to clearly understand which clinical indicators or neural features contributed to the output. Explainable AI (XAI) tools such as guided backpropagation~\cite{hu2021interpretable} and Shapley value-based attribution~\cite{chen2023algorithms} have improved explainability, but they remain fundamentally correlation-based and often yield inconsistent results across methods. Moving toward causal autoencoders~\cite{li2023causal} may represent an exploratory direction for establishing clearer and more robust causal relationships between brain activity and behavior or clinical decision-making processes.

\subsection{Generalization and personalization}  
From theoretical development to practical application, surrogate brain models must address two fundamental challenges: achieving robust \textbf{generalizability} under limited data conditions and advancing toward \textbf{personalization}. 

In brain dynamics modeling, neural activity generates high-dimensional state variables, \re{but} the scale of available data, particularly individual-specific data, remains limited, often leading to overfitting~\cite{libbrecht2015machine}. Future progress may proceed \re{in} two complementary directions. On one hand, enhancing data availability through international research collaborations and collecting dynamic data across diverse contexts, populations, and pathological states~\cite{allen2014uk} can provide richer and more comprehensive training resources. On the other hand, in data-constrained settings, incorporating neuroscientific priors and structured regularization into model design~\cite{keller2017predicting} may effectively enhance generalization performance and mitigate overfitting.

Nevertheless, robust generalization alone cannot fully address individual variability. The uniqueness of individual connectomes and the heterogeneity of regional brain dynamics call for models with strong adaptive capacity. Promising directions include adaptive transfer learning and few-shot fine-tuning~\cite{ding2023parameter}, which enable population priors to be efficiently adapted and calibrated with limited individual data. One promising example is the alignment of latent dynamics that ensures stabilization throughout the day in the BCI closed-loop center-out cursor control task~\cite{karpowicz2025stabilizing}. These strategies will help drive surrogate brain models from population-level generalizations toward individual-level customization, laying the foundation for personalized neuromodulation and precision interventions in neurological disorders.

\subsection{Ethical considerations}
Beyond technical advances, the development of surrogate brain models must also confront critical ethical challenges. These include protecting privacy and ensuring data security, addressing risks of algorithmic bias, and maintaining fairness and transparency when models are applied across diverse populations~\cite{kellmeyer2019artificial, zhui2024ethical}. Because surrogate brain models inherently rely on sensitive neural and clinical data, robust anonymization, secure data sharing frameworks, and rigorous governance structures are essential safeguards against misuse and unauthorized access. Furthermore, surrogate brain models may inadvertently inherit and amplify biases present in the training data~\cite{zhui2024ethical}. Applying biased models to different patient populations could lead to discriminatory or erroneous outcomes in clinical practice. Proactively addressing these ethical concerns is therefore essential for ensuring that surrogate brain models are developed and deployed responsibly, ethically, and equitably.

Although the challenges discussed above remain substantial, including limited generalizability under scarce data conditions, the need \re{to} balancing representational capacity with interpretability, and the importance of addressing ethical considerations, several promising directions for overcoming these obstacles have been outlined. These advances pave the way toward individualized surrogate brain models with enhanced accuracy, increased robustness, and broader applicability.
Through continued interdisciplinary collaboration, surrogate brain models may not only deepen our understanding of brain dynamics but also transform neurological care, ultimately enabling precision medicine and promoting equitable brain health.

\section*{ACKONWLEDGE}
We would like to thank Dr. Rui Liu, Dr. Haiyan Wu, Dr. Chaoming Wang, Mr. Weibin Li, Mr. Kaining Peng for their discussions and suggestions. 

\section*{FUNDING}
This work was supported in part by the National Natural Science Foundation of China (62472206, 12401233, 12141107), the National Key R\&D Program of China (2021ZD0201300), the NSFC International Creative Research Team (W2541005), the GuangDong Basic and Applied Basic Research Foundation (2025A1515011645), the Guangdong-Dongguan Joint Research Fund (2023A1515140016),   the Shenzhen Science and Technology Innovation Committee (RCYX20231211090405003, KJZD20230923115221044), the Science and Technology Commission of Shanghai Municipality (24Y22800200), Cross Disciplinary Research Team on Data Science and Intelligent Medicine (2023KCXTD054), Guangdong Provincial Key Laboratory of Mathematical and Neural Dynamical Systems (2024B1212010004), and the open research fund of the Center for Computational Science and Engineering at Southern University of Science and Technology.
\bibliography{ref} 
% \bibliography{ref}% common bib file

\end{document}